\documentclass[conference]{IEEEtran}

\usepackage{amssymb}
\usepackage{amsmath}
\usepackage{graphicx}
\usepackage{setspace}
\usepackage{color}
\usepackage{balance}

\begin{document}
%
\title{High-Speed Visible Light Indoor Networks\\ Based on Optical Orthogonal Codes and Combinatorial Designs}

\author{\IEEEauthorblockN{Mohammad Noshad and Ma\"{\i}t\'{e} Brandt-Pearce}\\
\IEEEauthorblockA{Charles L. Brown Department of Electrical and Computer Engineering\\
University of Virginia\\ Charlottesville, VA 22904\\
Email: mn2ne@virginia.edu, mb-p@virginia.edu}
}

\markboth{}{Shell \MakeLowercase{\textit{et al.}}: Bare Demo of IEEEtran.cls for Journals}

\maketitle

\begin{abstract}
Interconnecting devices in an indoor environment using the illumination system and white light emitting diodes (LED) requires adaptive networking techniques that can provide network access for multiple users. Two techniques based on multilevel signaling and optical orthogonal codes (OOC) are explored in this paper in order to provide simultaneous multiple access in an indoor multiuser network. Balanced incomplete block designs (BIBD) are used to construct multilevel symbols for $M$-ary signaling. Using these multilevel symbols we are able to control the optical peak to average power ratio (PAPR) in the system, and hereby control the dimming level. In the first technique, the $M$-ary data of each user is first encoded using the OOC codeword that is assigned to that user, and then it is fed into a BIBD encoder to generate a multilevel signal. The second multiple access method uses sub-sets of a BIBD code to apply multilevel expurgated pulse-position modulation (MEPPM) \cite{Multilevel-EPPM12} to the data of each user. While the first approach has a larger Hamming distance between the symbols of each user, the latter can provide higher bit-rates for users in VLC systems with bandwidth-limited LEDs.
\end{abstract}


\section{Introduction}
\PARstart{V}{isible} light communications (VLC) are considered a strong contender for the next generation of indoor communications and networking \cite{Indoor-OWC-Haas11}. VLC's immunity to radio-frequency (RF) interference, low impact on human health, and high data-rate capacity have made it an appealing technology to provide high-speed access for tablets, phones, laptops and other devices in many indoor spaces.
The application of VLC to indoor environments, such as offices, houses, airplanes, hospitals and convention centers, requires the capability to provide simultaneous connection for a large number of users.
In this paper we introduce two techniques to provide high-speed multiple access for simultaneous users in a VLC system.

Integrating VLC networks with illumination systems imposes limitations on the modulations and networking techniques that can be used. White light emitting diodes (LED) are the most common optical sources that are used in VLC systems, and modulation schemes that can be used with these devices are limited. Because of the structure of these LEDs and their inherent nonlinearity, implementing modulation and networking approaches that require spectral-encoding is expensive and complicated. Therefore, time-spreading modulations, specially pulsed techniques, are the preferred modulation technique in LED-based VLC systems. Dimming is an important feature of indoor lighting systems through which the illumination level can be controlled. Including dimming in the VLC system requires further constraints on the networking schemes that can be used.
A practical VLC network should support various optical peak to average power ratios (PAPR) so that, for a fixed peak power, the average power, which is proportional to the illumination, can be regulated.

Optical code division multiple access (OCDMA) is a networking technique that provides multiple access by assigning binary signature patterns to users \cite{OCDMA}. Among various OCDMA forms that have been proposed, time-spreading OCDMA is of most interest for indoor VLC system, since it can simply be implemented by turning the LEDs on and off \cite{OWC-OCDMA12}. In this type of OCDMA, binary sequences with special cross-correlation constraints, so-called optical orthogonal codes (OOC), are used to encode the data of users in the time-domain \cite{OCDMA1-89}. Codewords of an OOC are binary sequences that meet a given correlation constraint \cite{OOC-89}. The application of OOCs to VLC networks requires codes with a wide range of parameters for different dimming levels, which may not be practical for a network with a fixed number of users.

In \cite{VLC-JLT13} we proposed to apply expurgated pulse-position modulation (EPPM) \cite{EPPM12} and multilevel-EPPM (MEPPM) \cite{Multilevel-EPPM12} to indoor VLC systems in which the dimming can be done by simply changing the generating balanced incomplete block design (BIBD) code.
%
%
%
In this work, two networking methods are proposed based on MEPPM to provide multiple access for simultaneous users in a VLC system. These two techniques, which can be considered as synchronous OCDMA methods, enable users in a VLC network to have high-speed access to the network. In the first method we assign one OOC codeword to each user in order to encode its $M$-ary data and transmit the cyclic shifts of the assigned OOC codeword as symbols. For each user, every bit of this encoded binary sequence is multiplied by a BIBD codeword, and then the OOC-encoded BIBD codewords are added to generate a multilevel signal. Hence, the PAPR of the transmitted data can be controlled by changing the code-length to code-weight ratio of the BIBD code. In the second technique, a subset of BIBD codewords is assigned to each user, and then the MEPPM scheme is used to generate multi-level symbols using the assigned codewords. In this approach, users can have different bit-rates by partitioning the BIBD code into unequal-size subsets.

The organization of the rest of the paper is as follows. In Section II, we describe the indoor VLC network. The two networking methods to provide simultaneous access for different users are proposed in Section III. The performance of the proposed techniques are compared using numerical results in Section IV. Finally, Section V concludes the paper.

\section{System Description}

\begin{figure} [!t]
    \begin{center}
    \scalebox{0.32}{\includegraphics{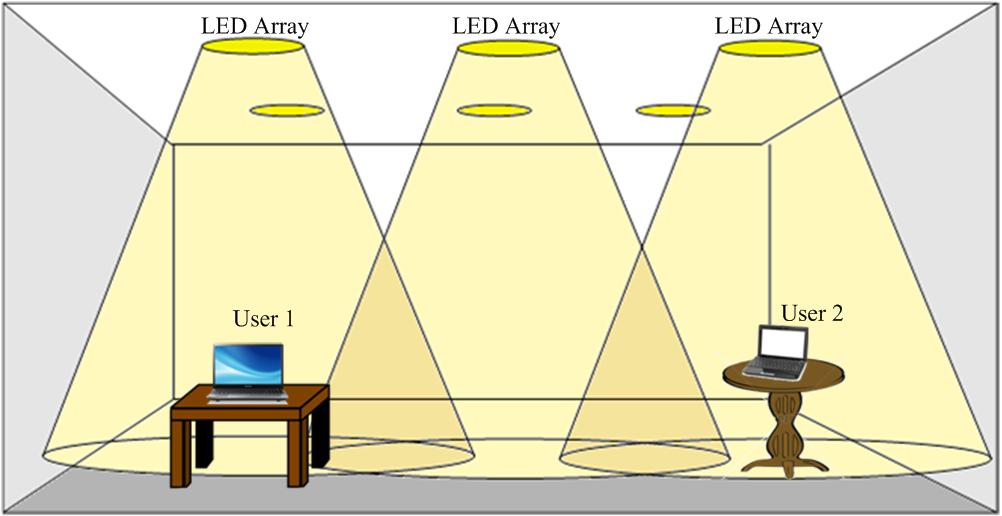}}
    \end{center}
    \vspace*{0.0 in}
    \caption{The configuration of a VLC network with LED arrays and two users.}
    \label{VLC}
    \vspace*{-0.0 in}
    \end{figure}

In LED-based VLC systems, both lighting and communications needs can be addressed at the same time. Due to eye-safety regulations and low cost, white LEDs are preferred over lasers for indoor communication applications \cite{Indoor-OWC-Haas11}. The downlink configuration of a VLC system is shown in Fig.~\ref{VLC}, where arrays of white LEDs are used as sources. Here, we consider LED arrays as access points, and we assume that all LED arrays in a room are synchronous and transmit the same data.
For each user, usually the strongest received power is the one that corresponds to the signal received from the direct path of the closest LED array, and therefore, it is considered as the main signal and the main data source. In situations where the direct path to the main source is blocked, the data can be retrieved using the multipath signals received from the non-line-of-sight (NLOS) paths \cite{VLC-JLT13}.
In this paper, VLC is assumed to be used only for the downlink, and infrared (IR) communications is used for the uplink channel to avoid self-interference from the full-duplex communication. 

In this work, BIBDs are used to modulate to LEDs. We describe a BIBD code \cite{Code-Designs} with parameters ($Q$,$K$,$\lambda$), where $Q$, $K$ and $\lambda$ respectively denote the code-length, the Hamming weight and the cross-correlation between any two codewords. The $m$th codeword is the binary vector $\mathbf{c}_m=(c_{m1}, c_{m2},\dots, c_{mQ} )$, $m = 1, 2, \dots, M $. The following relation holds between $\mathbf{c}_{m}$'s and is referred to as the fixed cross-correlation property \cite{Noshad11}:
    \begin{align}\label{Cross Correlation Property}
        \sum_{i=1}^{F} c_{mi} c_{ni} = \left\{ \begin{array}{lll}
                                        K            &; \, m = n, \\
                                        \lambda      &; \, m \neq n
                                        \end{array}
              \right.  .
    \end{align}

In \cite{Multilevel-EPPM12}, multilevel signals are constructed by combining multiple BIBD codes, which are then used as modulation symbols in multilevel EPPM (MEPPM). Each symbol-time is divided into $Q$ equal time-slots, and then each LED is turned on and off according to a BIBD codeword. Four different MEPPM schemes are introduced in \cite{Multilevel-EPPM12}, two of which (called type-I and type-II MEPPM) have a PAPR of $Q/K$ and are used in this paper.

\begin{figure} [!t]
    \begin{center}
    \scalebox{0.24}{\includegraphics{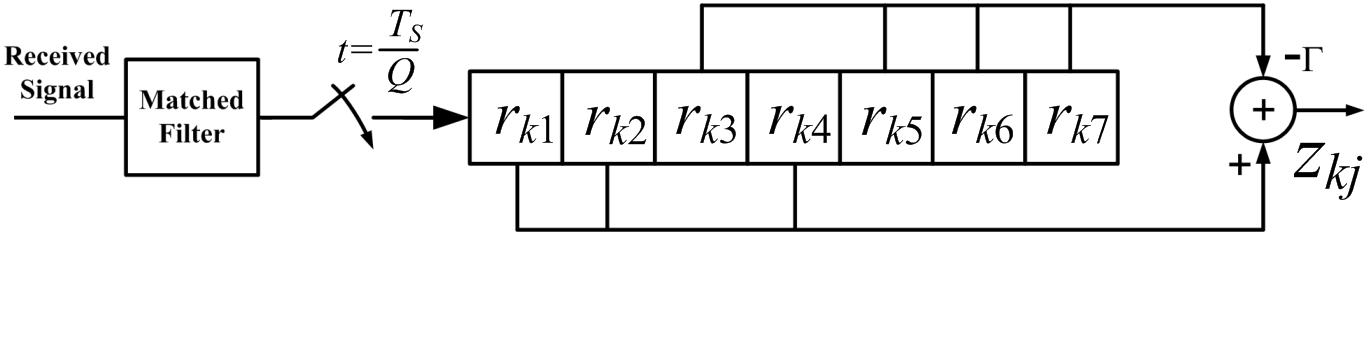}}
    \end{center}
    \vspace*{0.0 in}
    \caption{Receiver for MEPPM using a (7,3,1)-BIBD code.} 
    \label{EPPM Decoder}
    \vspace*{-0.0 in}
    \end{figure}

The structure of a simple receiver for EPPM and MEPPM using a shift-register is shown in Fig.~\ref{EPPM Decoder}. In this receiver, for symbol-epoch $k$, the sampled data at the output of a pulse-matched filter, $\mathbf{r_k}$, is stored in a shift register, and then is circulated inside it to generate vector $\mathbf{z}_k=(z_{k1},z_{k2},\dots,z_{kQ})$, $z_{kj}=\langle  \mathbf{r}_k ,\mathbf{c}_j \rangle$, at the output of the differential circuit, where $\langle \mathbf{x}, \mathbf{y} \rangle$ denotes the dot product of the vectors $\mathbf{x}$ and $\mathbf{y}$.
In this figure, $T_s$ is the symbol time and $\Gamma=\lambda/(w-\lambda)$. The wires of the lower branch are matched to the first codeword of the BIBD code, $\mathbf{c}_1$, and those of the upper branches are matched to its complement. This receiver is equivalent to the correlation decoder, which is shown in \cite{EPPM12} to be the optimum decoder for additive-white-Gaussian-noise (AWGN) channels.

\section{Multiple Access for Indoor Optical Networks}
Indoor optical networks must be able to provide simultaneous access for multiple users. Recently multi-input multi-output (MIMO) techniques, repetition coding, spatial multiplexing and spatial modulation, have been compared in \cite{OWC-MIMO13}.
Optical code division multiple access (OCDMA) can instead be used to fulfill this need. The two main techniques for implementing the codes in OCDMA networks are spectral encoding and time spreading codes \cite{OCDMA}. For indoor communication purposes, time spreading codes are preferred over spectral encoding since they have lower implementation cost. In VLC systems, since the access points are illumination sources and are the same for all users, synchronous OCDMA techniques can be used for the downlink. In \cite{OWC-OCDMA12}, synchronous time-spreading OCDMA using optical orthogonal codes (OOC) and on-off keying (OOK) was studied. For this case, because of the bandwidth limit of LEDs and the long length of OOCs needed, the data-rate for each user is limited. An efficient technique to increase the data-rate in OOC-based OCDMA systems is $M$-ary modulation using cyclic shifts of the OOC codewords, which is called code cycle modulation (OOC) in \cite{OCDMA-M-ary-06}. In this modulation, any cyclic shift of an OOC codeword with length $L$ is considered as a symbol, and therefore, the bit-rate is increased by a factor of $\log_2 L$. This technique requires perfect synchronization between transmitter and receiver.

A limitation of using OOCs in VLC systems is their incompatibility with the dimming feature. For a OOC OCDMA system that uses an OOC with length $L$ and weight $w$, the PAPR is $L/w$. Furthermore, for an OOC code with length $L$, weight $w$, and cross-correlation $\alpha$, the number of codewords, $N$, is bounded by the Johnson bound \cite{OOC-89}
    \begin{align}\label{Johnson Bound}
        N \leq \Bigg\lfloor \frac{1}{w} \bigg\lfloor \frac{L-1}{w-1} \Big\lfloor \frac{L-2}{w-2} \dots \Big\lfloor \frac{L-\alpha}{w-\alpha} \Big\rfloor \dots \Big\rfloor  \bigg\rfloor \Bigg\rfloor.
    \end{align}
Changing the pulse duty-cycle is not possible for bandwidth-limited sources, while changing the pulse amplitude requires a complex tuner circuit due to the source nonlinearity.
Therefore, in an OCDMA network with a given number of users, changing the PAPR requires employing a new OOC code, which, considering (\ref{Johnson Bound}), may not be possible for high PAPRs.

In this section, two networking methods based on MEPPM are introduced in order to not only provide multiple access for different users in an indoor VLC network, but also provide $M$-ary transmission for each user so that a higher data-rate can be achieved.

    \begin{figure} [!t]
    \vspace*{-0.0 in}
    \begin{center}
    \scalebox{0.25}{\includegraphics{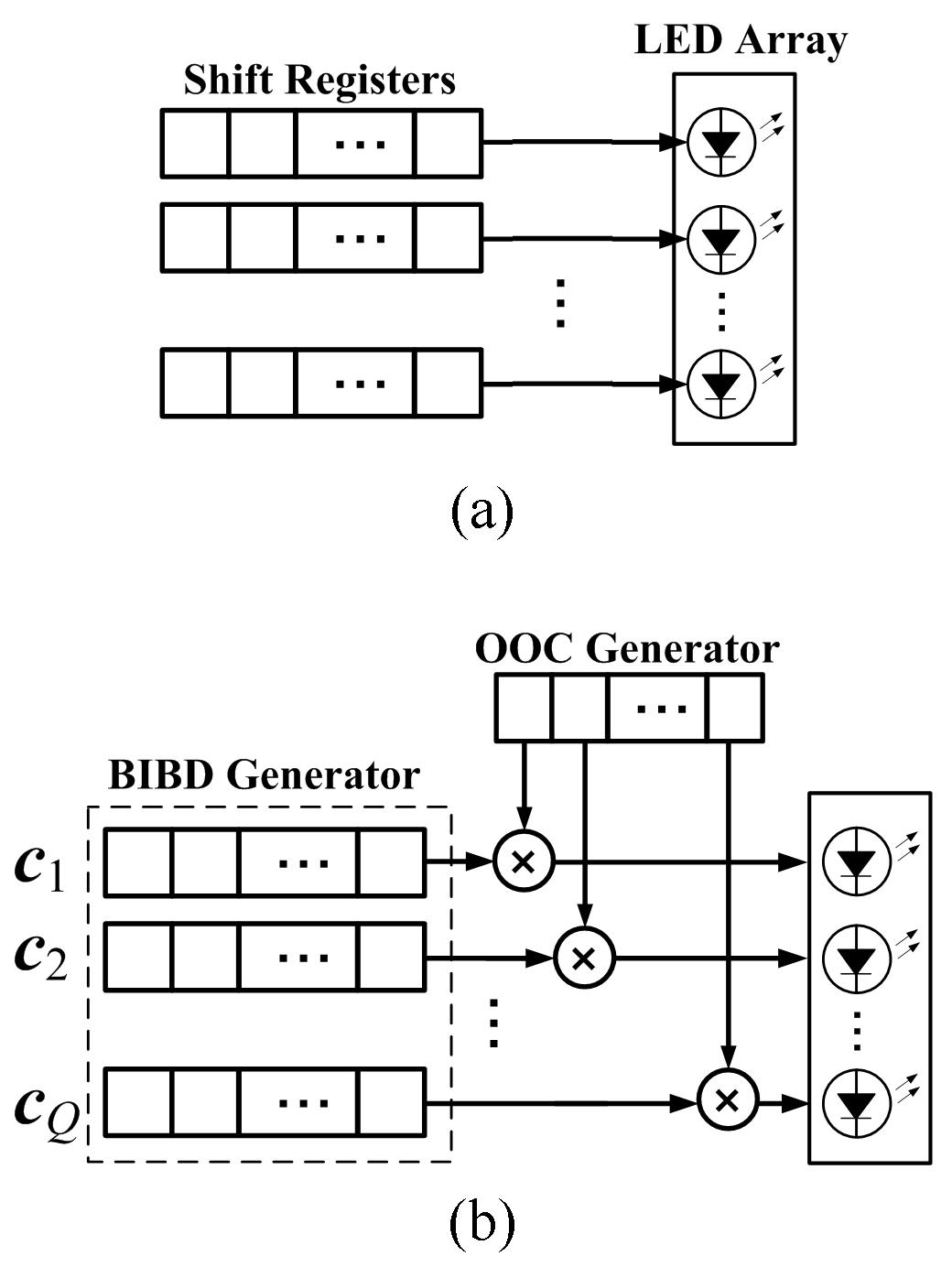}}
    \end{center}
    \vspace*{-0.0 in}
    \caption{Transmitter structure and symbol generation using shift registers for (a) MEPPM, and (b) coded-MEPPM.}
    \label{C-MEPPM Transmitter}
    \end{figure}
    \begin{figure} [!b]
    \vspace*{-0.0 in}
    \begin{center}
    \scalebox{0.20}{\includegraphics{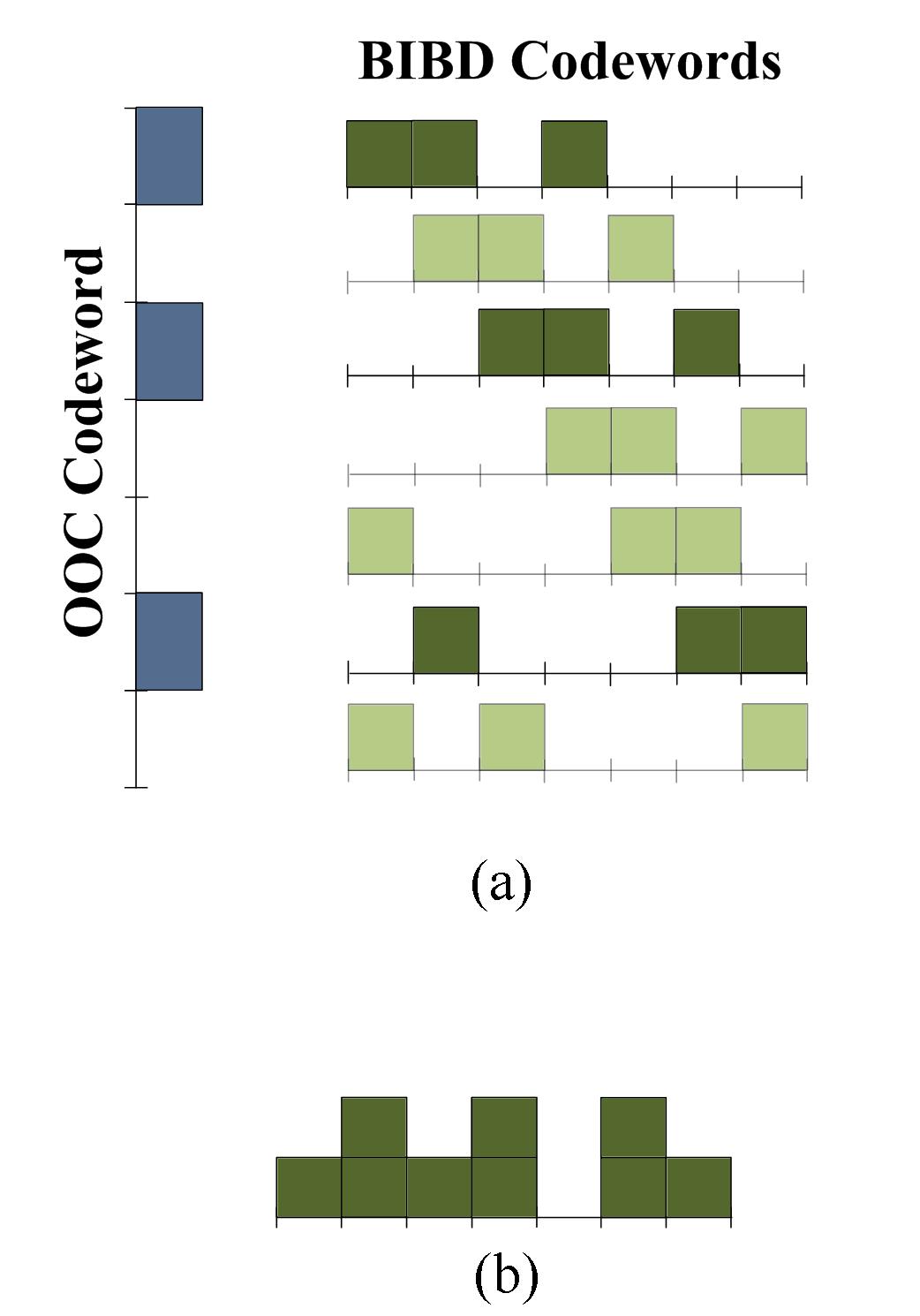}}
    \end{center}
    \vspace*{-0.0 in}
    \caption{(a) Symbol generation for coded-MEPPM using the (1010010) OOC codeword and a (7,3,1)-BIBD, and (b) resulting symbol.}
    \label{C-MEPPM Symbol}
    \end{figure}

\subsection{Networking Using Coded-MEPPM}
OOCs are used in OCDMA networks to provide simultaneous access for users by assigning signature pattern to each user \cite{OOC-89}. These codes are usually implemented in the time domain and are used as time-spreading codes. In our proposed technique that we call coded-MEPPM (C-MEPPM), we combine OOCs with MEPPM to provide multi access and high data-rate for each user. Unlike traditional OCDMA systems, in our approach the OOC codewords are implemented in the code domain, and are applied on the codewords of a BIBD code in code-space. Let $\mathbf{d}_n=[d_{n1},d_{n2},\dots,d_{nL}]$, $n=1,2,\dots,N$, $d_{n\ell} \in \{0,1\}$, be the $n$th codeword of an OOC with length $L$, weight $w$, cross-correlation $\alpha$, and $N$ codewords. By assigning the $n$th OOC codeword to user $n$, its transmitted signal for the $m$ symbol in the $M$-ary constellation is given by
    \begin{align}\label{coded-MEPPM Symbols}
        \mathbf{u}_{m,n} = \frac{1}{Q} \sum_{\ell=1}^L d_{n \ell} \, \mathbf{c}{^{(m)}_{\ell}}.
    \end{align}
where the notation $\mathbf{x}^{(m)}$ is the $m$th cyclic shift of the vector $\mathbf{x}$.
In this manner, the symbols of user $n$ are cyclic shifts. From (\ref{coded-MEPPM Symbols}), the length of the OOC should be no longer than the number of BIBD codewords, i.e., $L \leq Q$. In this work, for a fixed $Q$ we use an OOC with $L=Q$, since the larger the OOC-length, the higher performance it can achieve. Fig.~\ref{C-MEPPM Transmitter} shows the transmitter structure using shift-registers for coded-MEPPM, and Fig.~\ref{C-MEPPM Symbol} shows the generated multilevel symbol using the (1010010) OOC codeword and a (7,3,1) BIBD code with $\mathbf{c}_1 = (1101000)$.

At the receiver side, the correlation-receiver in Fig.~\ref{EPPM Decoder} is used to decode the BIBD codewords. Let $\mathbf{s}_{k,n}$, $n\in \{1,2,\dots,N\}$, be the transmitted vector for user $n$ in symbol-time $k$. When user $n$ is the only active user and is transmitting symbol $m_n$, the mean value of the output of the correlator at symbol-time $k$ can be written as
    \begin{align}\label{C-MEPPM output}
        E\Big\{ z_{kj} \Big| \mathbf{s}_{k,n}&=\mathbf{u}_{m_n,n} \Big\} \nonumber\\
        &= \Lambda_0 \bigg( \Big\langle\mathbf{u}_{m_n,n},\mathbf{c}_j\Big\rangle -\Gamma\Big\langle\mathbf{u}_{m_n,n},\overline{\mathbf{c}_j}\Big\rangle  \bigg) \nonumber\\
        &= \Lambda_0 (K/Q) d_{n (j-m_n)}.
    \end{align}
where $\Lambda_0=\eta \frac{P_0}{h \upsilon} \frac{\log_2 Q T_b}{Q}$, $P_0$ is the peak received power, $h$ is Planck's constant, $\upsilon$ is the central optical frequency, and $\eta$ is the efficiency of the photodetector.
The vector form of (\ref{C-MEPPM output}) can be written as
    \begin{align}\label{C-MEPPM vector output}
        E\Big\{ \mathbf{z}_{k} \Big| \mathbf{s}_{k,n}&=\mathbf{u}_{m_n,n} \Big\}
        = \Lambda_0 (K/Q) \mathbf{d}{_n^{(-m_n)}}.
    \end{align}

\begin{figure} [!t]
    \begin{center}
    \scalebox{0.26}{\includegraphics{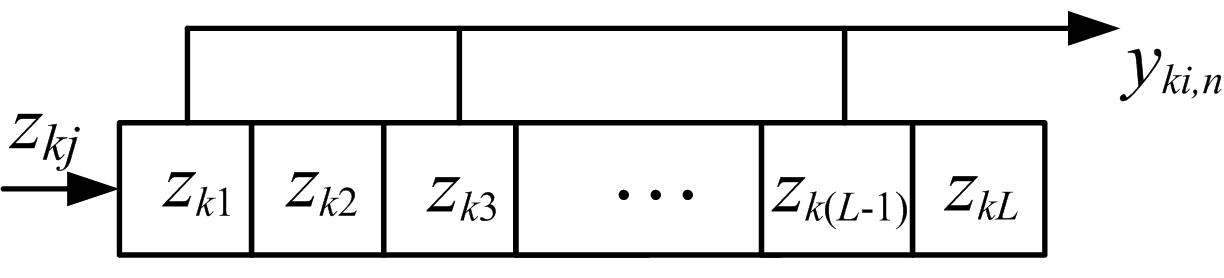}}
    \end{center}
    \vspace*{-0.0 in}
    \caption{OOC decoder using a shift register.}
    \label{OOC Decoder}
    \vspace*{-0.0 in}
    \end{figure}

\begin{figure} [!b]
    \vspace*{-0.0 in}
    \begin{center}
    \scalebox{0.51}{\includegraphics{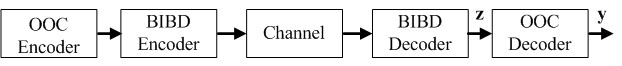}}
    \end{center}
    \vspace*{-0.0 in}
    \caption{Schematic view of a code-MEPPM system using OOC.}
    \label{OOC-BIBD}
    \end{figure}

Thus the output of the correlation decoder resembles the received signal of an OOC-OCDMA system using OOC. A correlation decoder using a shift-register can be used to decode the OOC code \cite{OCDMA1-89}, as shown in Fig.~\ref{OOC Decoder}. The overall system using OOC and MEPPM encoders and decoders is depicted in Fig.~\ref{OOC-BIBD}. The expected value of the $i$th output of the OOC-correlator that matches the OOC-codeword $\mathbf{d}_n$ in symbol-time $k$, $y_{ki,n}$ $i=1,2,\dots,L$, can be written as
    \begin{align}\label{}
        E\Big\{ y_{ki,n} \Big| \mathbf{s}_{k,n}=\mathbf{u}_{m_n,n} \Big\}=\left\{ \begin{array}{lll}
                                        \Lambda_0(K/Q) w            &; \, i = m_n, \\
                                        \leq \Lambda_0 (K/Q) \alpha      &; \, i \neq m_n.
                                        \end{array}
              \right.
    \end{align}

In the general case when all users are active, given that symbol $m_n$ is sent by user $n$, for $n=1,2,\dots,N$, in symbol-time $k$, the mean value of $\mathbf{z}_k$ is
    \begin{align}\label{}
        E\Big\{\mathbf{z}_k \Big| \mathbf{s}_{k,n'}=\mathbf{u}_{m_{n'},n'}, n'=1,2,&\dots,N \Big\} \nonumber\\&= \Lambda_0 \Big(\frac{K}{Q}\Big) \sum_{n'=1}^N \mathbf{d}{_{n'}^{(-m_{n'})}},
    \end{align}
and, accordingly, the output of the OOC-correlator of user $n$ is
    \begin{align}\label{}
        E&\Big\{y_{ki,n} \Big| \mathbf{s}_{k,n'}=\mathbf{u}_{m_{n'},n'}, n'=1,2,\dots,N \Big\} \nonumber\\
        &= \Lambda_0 \Big(\frac{K}{Q}\Big) \sum_{n'=1}^N \langle \mathbf{d}{_{n'}^{(-m_{n'})}}, \mathbf{d}{_{n}^{(i)}} \rangle, \, \, i=1,2,\dots,L \nonumber\\
        &=\Lambda_0 \Big(\frac{K}{Q}\Big) \langle \mathbf{d}{_{n}^{(-m_n)}}, \mathbf{d}{_{n}^{(i)}} \rangle + \underbrace{\Lambda_0 \Big(\frac{K}{Q}\Big) \sum_{\substack{{n'=1} \\ n' \neq n}}^N \langle \mathbf{d}{_{n'}^{(-m_{n'})}}, \mathbf{d}{_{n}^{(i)}} \rangle}_{MAI} .
    \end{align}
The second term above can be considered as the multiple access interference (MAI), which is caused by the signals from other users.

Let us compare the proposed coded-MEPPM to OOC-OCDMA with OOK.
For an OOC-OCDMA VLC network with $N$ active users, since the LED arrays transmit the signal of all users, $N$ LEDs are needed. Therefore, assuming $P_0$ to be the total power received from one LED array, the data of each user is received at a power level of $P_0/N$. For an OOC with length $L=Q$, weight $w$, and cross-correlation $\alpha$ used in a OOC-OCDMA network, the Euclidean distance between bit "0" and "1" is $d_{\textmd{OOC}}=\Lambda_0 \sqrt{w} /(\log_2 Q N)$. While for a coded-MEPPM, because of the fixed-cross correlation of $(Q,K,\lambda)$-BIBD codewords, at most $K$ LEDs are needed. So, the distance between the symbols is given by $d_{\textmd{C-MEPPM}}=(\Lambda_0/Q) \sqrt{2(K-\lambda)(w-\alpha)}$. Comparing these two distances, we can see that $d_{\textmd{C-MEPPM}} = d_{\textmd{OOC}} (N \log_2 Q/Q)\sqrt{(K-\lambda)\frac{2(w-\alpha)}{w}}$, and in networks for which $N>\frac{Q}{\log_2 Q \sqrt{(K-\lambda)}}$, the coded-MEPPM technique has a larger minimum distance between symbols compared to OOC-OCDMA.

\subsection{Networking Using Divided-MEPPM}
In the second proposed technique, which we call divided-MEPPM (D-MEPPM), the generating BIBD code is divided into several smaller codes, and a different set of codes is assigned to each user. Then, similar to the MEPPM scheme, each user uses its codeword set to generate multilevel symbols. Let $q_n$ be the number of BIBD codewords that are assigned to user $n$, such that $q_1+q_2+\dots+q_N=Q$, and let $\mathcal{C}_n$, $\Big|\mathcal{C}_n\Big|=q_n$, be the set of codewords that are assigned to user $n$, such that $\mathcal{C}_n \cap \mathcal{C}_m = \emptyset$ for any $n \neq m$, and $\mathcal{C}_1 \cup \mathcal{C}_2 \cup \dots \cup \mathcal{C}_N = \{\mathbf{c}_1,\mathbf{c}_2,\dots,\mathbf{c}_Q\} $. This can be considered as a kind of CDMA, where distinct codeword sets with cross-correlation $\lambda$ are assigned to users. Using this definition, user $n$ can utilize an $\ell_n$-branch MEPPM, $1 \leq \ell_n < q_n$ , for $M$-ary transmission using $\mathcal{C}_n$. It can use either MEPPM type-I or type-II, yielding a constellation of size ${q_n}\choose{\ell_n}$ for type-I and ${q_n+\ell_n}\choose{\ell_n}$ for type-II MEPPM \cite{Multilevel-EPPM12}. The symbol $m$ of user $n$ can be expressed as
    \begin{align}\label{}
        \mathbf{u}_{m,n} = \frac{1}{Q} \sum_{\ell=1}^{\ell_n} \mathbf{c}_{j_{\ell,m}},
    \end{align}
where $\mathbf{c}_{j_{\ell,m}} \in \mathcal{C}_n$ for $\ell=1,2,\dots,\ell_n$. The smallest distance between two symbols of user $n$ is $\Lambda_0 2(K-\lambda)/w$. Each user can generate these symbols using the transmitter shown in Fig.~\ref{C-MEPPM Transmitter}-(a).

An advantage of this networking technique over the one presented in Section III-A is its potential to provide different data-rates for different users, which can be done by assigning unequal size subsets to users. Thus, we provide a larger number of BIBD codewords to users in the network requiring a higher bit-rate. 

    \begin{figure} [!t]
    \vspace*{0.0 in}
    \begin{center}
    \scalebox{0.47}{\includegraphics{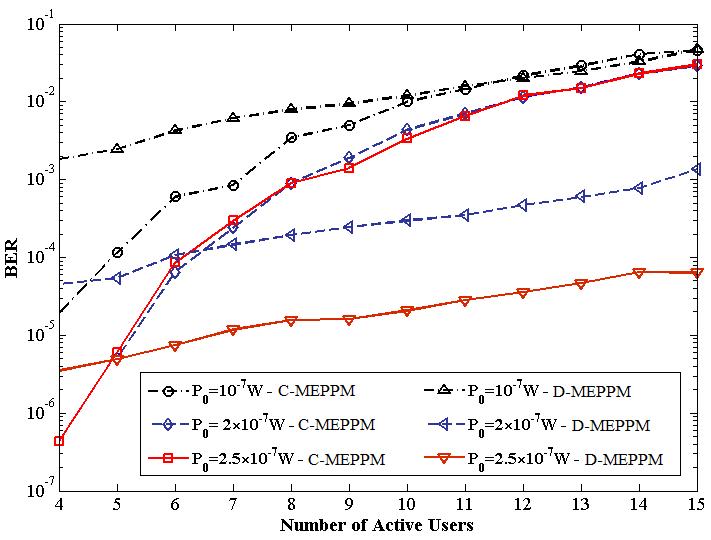}}
    \end{center}
    \vspace*{-0.0 in}
    \caption{Simulated BER versus the number of active interfering users for coded-MEPPM (C-MEPPM) and divided-MEPPM (D-MEPPM) with different peak power levels for $\eta=0.8$, central wavelength $\lambda=650$ and background power $P_b=0.1$ $\mu W$.}
    \label{OWN}
    \end{figure}
    \begin{figure} [!b]
    \vspace*{-0.0 in}
    \begin{center}
    \scalebox{0.48}{\includegraphics{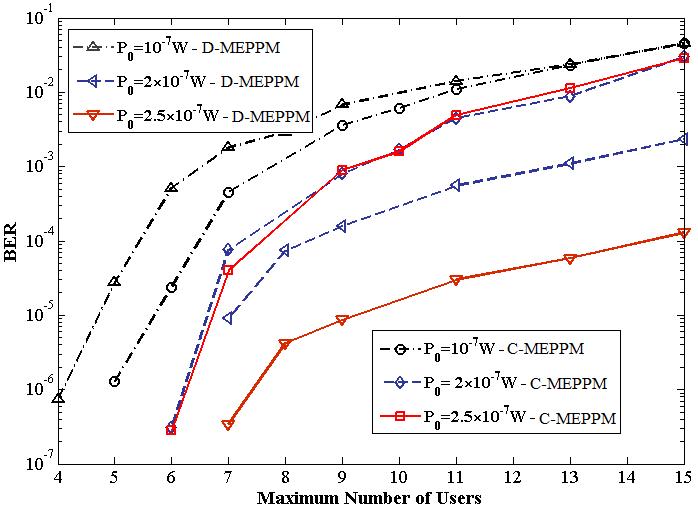}}
    \end{center}
    \vspace*{-0.0 in}
    \caption{Simulated BER versus the number of active interfering users for coded-MEPPM and divided-MEPPM with different peak power levels.}
    \label{OWN-2}
    \end{figure}

\section{Numerical Results}
A comparison between the performance of the coded-MEPPM and divided-MEPPM using the same BIBD code is shown in Fig.~\ref{OWN}. In this figure the BER is plotted versus the number of active interfering users for three different peak power levels. For the coded-MEPPM, a (67,33,16)-BIBD code is used to construct the symbols, and a (67,13,3)-OOC is used to encode the users' data. For this system, each user has 67 symbols, and the maximum number of users is 16. The Euclidean distance between two symbols of a user is $\Lambda_0\sqrt{340}/67$. In divided-MEPPM, distinct sets of 4 BIBD codewords are assigned to users, and therefore, the maximum number of users is again 16. For this technique, each user utilizes type-II MEPPM and has 70 symbols. The minimum Euclidean distance between two symbols is $\Lambda_0 \sqrt{34}/67$. According to these results, for weak peak powers the error probability of coded-MEPPM is lower than divided-MEPPM since it has larger distance between its symbols and its performance is only limited by multiple access interference (MAI), while the performance of divided-MEPPM is limited by the shot noise.
    \begin{figure} [!b]
    \vspace*{-0.0 in}
    \begin{center}
    \scalebox{0.42}{\includegraphics{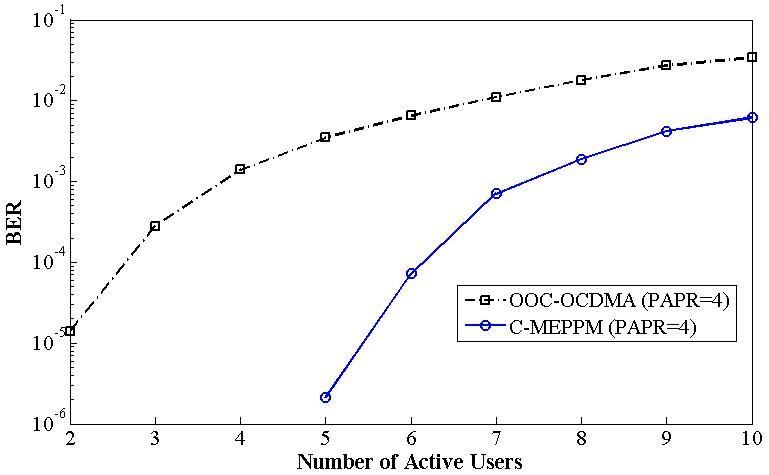}}
    \end{center}
    \vspace*{-0.0 in}
    \caption{Simulated BER versus the number of active interfering users for coded-MEPPM OOC-OCDMA for a PAPR of 4, peak received power $P_0=0.1$ $\mu W$ and background power $P_b=0.1$ $\mu W$.}
    \label{OWN-3}
    \end{figure}

Fig.~\ref{OWN-2} shows the BER of the coded-MEPPM and divided-MEPPM versus the maximum number of users for three different peak power levels. For coded-MEPPM results, different OOC codes are used to provide various maximum number of users. Similarly, for divided-MEPPM, the size of the sub-sets of BIBD codewords are chosen so that we achieve the maximum throughput for each point on the curve. The parameters are selected such that the number of symbols for all cases is approximately the same. According to these results, the coded-MEPPM performs better than divided-MEPPM only for low received peak power cases, and for strong received power levels the divided-MEPPM has a lower BER compared to coded-MEPPM.

The simulated BER of C-MEPPM is compared to an OCDMA network using OOK modulation and a (14,7,4)-OOC in Fig.~\ref{OWN-3} for a VLC network with maximum 10 users. The results are plotted versus the number of users, $N$. The C-MEPPM scheme is using a (101,25,6)-BIBD and a (101,11,2)-OOC to transmit the data at a 200 Mb/s bit-rate. The PAPR for both OOC-OCDMA and C-MEPPM systems is 4, and thus, they are providing the same illumination level. OOC-OCDMA is using almost the same length of time-chips as C-MEPPM in order to have the same bit-rate. Due to the large ratio of $\alpha/w$ of the OOC code, interfering users in OOC-OCDMA technique are causing largeer interference effect on the main user compared to C-MEPPM, and therefore, it has a larger error probability for the same number of active users.

\section{Conclusion}
Two multiple-access methods, are introduced and compared. The first method, using both OOC and BIBD codes to encode the user data, is shown to have a large distance between symbols. The second technique only uses BIBD codes to construct MEPPM symbols, and, therefore, can have a large constellation size and consequently high bit-rate for each user. According to the simulation results, for low SNR cases the coded-MEPPM technique achieves a lower BER compared to divided-MEPPM, while the latter is preferred in high SNR regimes since it has a lower MAI effect. They both outperform the conventional OOC-OCDMA.

\section{Acknowledgment}
This research was funded by the National Science Foundation (NSF) under grant number ECCS-0901682.

\bibliographystyle{IEEEtran}
\bibliography{EPPM}
\balance

\end{document}